\documentclass[aoas,MSNbibl,nameyear,dvips]{arximspdf}

\doi{10.1214/11-AOAS453A}
\volume{5}
\issue{2B}
\pubyear{2011}
\firstpage{1425}
\lastpage{1427}
\referstodoi{10.1214/10-AOAS453}

\makeatletter
\renewcommand{\epsilon}{\varepsilon}
\makeatother

\begin{document}
\begin{frontmatter}

\title{Discussion of ``Network routing in a~dynamic~environment''\thanksref{T2}}
\runtitle{Discussion}
\pdftitle{Discussion of Network routing in a dynamic environment by N.
D. Singpurwalla}

\begin{aug}
\author[A]{\fnms{Andrew C.} \snm{Thomas}\corref{}\ead[label=e1]{acthomas@stat.cmu.edu}}
\and
\author[A]{\fnms{Stephen E.} \snm{Fienberg}\ead[label=e2]{fienberg@stat.cmu.edu}}
\runauthor{A. C. Thomas and S. E. Fienberg}
\affiliation{Carnegie Mellon University}
\address[A]{Department of Statistics\\
Carnegie Mellon University\\
5000 Forbes Avenue \\ Baker Hall 132 \\ Pittsburgh,
Pennsylvania 15213\\USA\\
\printead{e1}\\
\hphantom{\textsc{E-mail:}} \printead*{e2}} 
\end{aug}
\thankstext{T2}{Supported in part by DARPA Grant 21845-1-1130102.}

\received{\smonth{1} \syear{2011}}



\end{frontmatter}

\section{Introduction}

An earlier version of Professor Singpurwalla's paper (which we refer to
as ``Singpurwalla'') has served as the springboard to our own
investigation of the issue of deployments of Improvised Explosive
Devices, or other obstacles with a large cost to overcome, which may be
placed stochastically, or by an adversarial agent, or both.

Rather than a decision-theoretic treatment, we consider a method based
in part on social network analytical methods, namely, that the
deployment pattern of IEDs induces a subgraph on a full road network,
and that the deployment on any given road is unknown to anyone
traversing the graph until arriving there, though there may be prior
information on the likelihood of a deployment.

The full treatment, as acknowledged in Singpurwalla, is illustrated by
\citet{thomas2011ecidglmictp}; here, we give a brief overview of our
method and how it compares with Singpurwalla's approach.

\section{Canadian traveler problems and network transition times}

Many analyses of social networks assume that the shortest path between
two individuals governs properties of their inter-relationship, and
this has led to many metrics constructed using geodesic distance to
approximate the importance of an individual [\citet{freeman1979csncc}].
If the streets were empty of traffic, a driver on the roads will think
the same way, taking the route that minimizes travel time. This is not
necessarily the case when the state of the roads is uncertain, such as
with traffic or construction, but is nicely encapsulated in the
``Canadian Traveler Problem'' formulation
[\citet{andreatta1988sspr}; \citet{schieber1991ctp};
\citet{papadimitriou1991spwm}]: a road may be impassable
because, with some probability, there is an obstacle that cannot be
traversed without waiting (in the eponymous case, a heavy snow fall).
If the probabilities are known in advance, but the actual states of the
roads are not known until reached, then an optimal route can be
calculated either through exact solution or simulation, by solving for
the distribution of travel times along any particular route, simulating
the blockages given their propensities.

Given that all roads have some probability of a blockage, IED
deployment or otherwise, we can evaluate a road's importance for travel
by comparing the average travel time if the road is active to that when
the road is blocked,\vadjust{\goodbreak} given a source and destination. The effective
difference in travel times is then a measure of the importance of the
road to that travel.

Figure 1 of Singpurwalla gives three potential paths between a source
$A$ and a destination (sink) $I$, along which the traveler may move. If
a road's state (in this case, a bridge's state) is discovered once one
of its connecting intersections is reached, then this will influence
the traveler as they move through the system. For this map, the
traveler would know immediately whether bridge 9 was traversable; the
only choice would then be if the route $\mathit{CDEI}$, or the route $\mathit{CDEFGHI}$,
are shorter than the direct route $\mathit{CI}$, though the risk remains. If
either of these routes is shorter when unblocked, it is the traveler's
decision to try the shorter route, with the risk of having to turn
back, or take the certain path without learning additional information.

\section{Additional covariates}

Singpurwalla's approach includes covariates in the likelihood for IED
deployments on particular stretches of road in the standard fashion of
a logistic regression. Examples include ``local'' characteristics like
the proximity to a center of commerce or worship, or the nature of the
road itself, such as its construction, capacity, and length, as well as
circumstances particular to the timing of any particular attack, like
the time of day or the weather.

For purely exploratory modeling, we can also consider the role of any
road in relation to the rest of the system of roads; for example, if
there are three parallel routes of equal length that a traveler can
take, the likelihood that any one of these routes will be blocked will
go down, all else being equal.

As we mentioned, social network analysts typically use measures of
``centrality'' on a graph to elicit information about the role of a
node or an edge on the network, often deriving these from the role of
shortest paths on the network. In this case, we can include the
importance of a road in the system by considering how the road affects
the Canadian Traveler: calculating the average additional travel that
would be necessary if the road were closed, rather than open, when the
traveler does not learn this until and unless they arrive at one of its
endpoints. Because this measures the importance of a road as a conduit
between two points, we have christened this quantity Canadian
Betweenness Centrality; whether or not it is calculated with the
possibility of other roads also being blocked is up to the analyst.

\section{Expert information}

Singpurwalla makes note of the encapsulation of prior information on
deployments on particular road systems according to the decision
maker's subjective or personal probability. Rather than including this
step directly into the probability of a deployment, we recommend a~%
slightly more roundabout approach: using the assessment of personal
probabilities by the decision maker or the experts to elicit a prior
distribution on the coefficients in the model [\citet
{garthwaite2005smfepd}], here corresponding to the $\beta$ terms in the
logistic regression.

Essentially, the modeler queries the expert about their estimated
probability of a deployment in a particular time period for all roads,\vadjust{\goodbreak}
then fits the model to estimate distributions on $\beta$ corresponding
to this uncertainty, having transformed the fractions between zero and
one into an unbounded region. In the notation of Singpurwalla, we set\vspace*{-1.5pt}
\[
\log\frac{p_l}{1-p_l} = \sum_{i=1}^k Z_{il}\beta_i + \epsilon_l,
\]
 where $\epsilon_l \sim N(0, \sigma^2)$. If we define $P =
[\log\frac{p_1}{1-p_1}\enskip\cdots\enskip \log\frac{p_n}{1-p_n}]'$, the
elicited prior distribution is
\[
\beta\sim N_k((Z'Z)^{-1}Z'P, (Z'Z)^{-1}\widehat{\sigma}^2).
\]
 If the estimate provided by the expert (or group of experts)
comes from a~bdistribution, we can carry out this procedure using draws
from this distribution by mixing across iterations.

\section{Conclusions}

Singpurwalla's approach has provided a crucial stimulus to our own
pursuit of the problem of road-blocking deployments. We all have a long
way to go before this framework can be applied practically. As noted by
Singpurwalla, the limited availability of this class of data makes it
difficult to validate any of our methods, especially the adversarial
nature. Even the data that are known to exist publicly, such as the
Wikileaks disclosure, are not in a form that makes our frameworks
applicable. We are therefore left to continue developing these ideas
through simulation and thought experiment until the time comes for
their more practical application.


%

\printaddresses

\end{document}